\documentstyle[12pt]{article}

\newcommand{\resection}[1]{\setcounter{equation}{0}\section{#1}}

\newcommand{\EQ}{\begin{equation}}
\newcommand{\EN}{\end{equation}}
\newcommand{\bea}{\begin{eqnarray}}
\newcommand{\eea}{\end{eqnarray}}

\newcommand{\be}{\beta}

\begin{document}
\setcounter{page}{0}
\topmargin 0pt
\oddsidemargin 5mm
\renewcommand{\thefootnote}{\fnsymbol{footnote}}
\newpage
\setcounter{page}{0}
\begin{titlepage}
\begin{flushright}
ISAS/60/94/EP
\end{flushright}
\vspace{0.5cm}
\begin{center}
{\large {\bf Sprectral Representation of Correlation Functions in
Two-dimensional Quantum Field Theories}} \\
\vspace{1.8cm}
{\large G. Mussardo}\\
\vspace{0.5cm}
{\em International School for Advanced Studies,\\
and \\
Istituto Nazionale di Fisica Nucleare\\
34014 Trieste, Italy}\\

\end{center}
\vspace{1.2cm}
\renewcommand{\thefootnote}{\arabic{footnote}}
\setcounter{footnote}{0}

\begin{abstract}
The non-perturbative mapping between different Quantum Field Theories and
other features of two-dimensional massive integrable models are discussed
by using the Form Factor approach. The computation of ultraviolet data
associated to the massive regime is illustrated by taking as an example the
scattering theory of the Ising Model with boundary.
\end{abstract}

\vspace{3cm}

Talk given at the {\em International Colloquium on Modern Quantum Field Theory
II} Tata Institute, Bombay, January 1994.
\end{titlepage}

\newpage

\resection{Introduction}

Many two-dimensional integrable statistical models with a finite
correlation length can be elegantly discussed in terms of relativistic
particles in bootstrap interaction \cite{Zam}. In this formulation, the
key object is the elastic $S$-matrix that describes the scattering processes
of the massive excitations in the Minkowski space. Once we know the exact
$S$-matrix of the model under analysis and the corresponding spectrum, we may
proceed further and compute several quantities of theoretical interest, among
them the central charge and the critical exponents of the conformal field
theory arising in the ultraviolet regime \cite{TBA}. Aim of this talk is to
discuss some features of the structure of integrable QFT in terms of the
properties of their correlation functions. The most promising method
for the computation of the correlation functions results to be the
{\em Form Factor Approach}, as originally proposed in \cite{Karowski,Smirnov}.
In the following I will try to point out the reasons of the successful
application of this approach together with several interesting properties
which come out as by-products of its theoretical formulation.
As a significant example of the computation of ultraviolet data in terms
of a resummation of the Form Factors, we will consider the exact
critical exponents of the energy and disorder operators in the Ising
model with boundary. The scattering theory for such system has been recently
proposed in \cite{GZ}.

\resection{Computation of Correlation Functions}

To fully appreciate the bootstrap approach to the computation
of the correlation functions, let us discuss the most common difficulties
which arise in the perturbative method. Let
\EQ
{\cal A}\,=\,{\cal A}_0\,+\,\lambda \,{\cal A}_{\rm int} \,\,\,
\EN
be the action of the theory, where ${\cal A}_0$ corresponds to a solvable
QFT (e.g. a free theory, CFT, etc.) whereas ${\cal A}_{\rm min}$ defines the
interactive part. For the perturbative definition of the Green functions
we have the formal expressions
\EQ
G^{(N)}(x_1,\ldots,x_N)\,=\,\sum_{k=0}\lambda^k \,G^{(N)}_k(x_1,\ldots,x_N)
\,\,\, ,
\EN
where $G^{(N)}_k(x_1,\ldots,x_N)$ is the $k$-th perturbative term.
The above expression usually suffers of several drawbacks:
\begin{itemize}
\item We may face, for instance, the renormalization problem, i.e. the
presence of infinities which should correctly be handled. In this case,
we need to express our final computation in terms of physical quantities by
means of the renormalization procedure. This is usually a painful task.
\item Assuming we were able to go through the renormalization
program, the resulting expression may present a low rate of convergence,
if any!
\end{itemize}
In the light of the above considerations, a more efficient way to compute
correlation functions is needed. This is provided by the spectral
representation methods, i.e. by the possibility to express the correlation
functions as an infinite series over multi-particle intermediate states. For
instance, in a QFT with a self-conjugate particle the two-point function
of an operator ${\cal O}(x)$ in real Euclidean space is given by
\begin{eqnarray}
& &\langle{\cal O}(x)\,{\cal O}(0)\rangle\,=
\sum_{n=0}^{\infty}
\int \frac{d\beta_1\ldots d\beta_n}{n! (2\pi)^n}
<0|{\cal O}(x)|\beta_1,\ldots,\beta_n>_{\rm in}{}_{\rm in}
<\beta_1,\ldots,\beta_n|{\cal O}(0)|0>
\label{correlation} \nonumber \\
& &\hspace{3mm} =\,\sum_{n=0}^{\infty}
\int \frac{d\beta_1\ldots d\beta_n}{n! (2\pi)^n}
\mid F_n(\beta_1\ldots \beta_n)\mid^2 \exp \left(-mr\sum_{i=1}^n\cosh\beta_i
\right)
\end{eqnarray}
where $r$ denotes the radial distance, i.e. $r=\sqrt{x_0^2 + x_1^2}$ and
$\beta$ the rapidity variable.
Similar expressions are obtained for higher point correlators. The functions
\EQ
F_n(\beta_1,\ldots,\beta_n)\,=\,<0|{\cal O}(0)|\beta_1,\ldots,\beta_n>
\EN
are the so-called {\em Form Factors}. Since the spectral representations
are based only on the completeness of the asymptotic states, they are general
expressions for any QFT. However, for integrable models, they become quite
effective because the exact computation of the form factors reduces to finding
a solution of a finite set of functional equations, as we will discuss below.
Other advantages of the spectral representation method for the correlation
functions may be summarized as follows:

\begin{enumerate}
\item It deals with physical quantities, i.e. there is no need of
renormalization and the coupling constant dependance is taken into account at
all orders by a closed expression of the Form Factors.
\item The above representation (\ref{correlation}) and similar expressions
for other correlators present a very fast rate of convergence for
all values of the scaling variable $(mr)$. This is quite expected for
large values of $(mr)$ which are dominated by the lowest massive state
but the surprising result is that in many cases the series is saturated
by the lowest terms also for small values of $(mr)$. This is
due to a threshold suppression phenomenon \cite{Camus} which we will
present below.
\item The two-point correlation functions (\ref{correlation}) have the form
of a Grand-Canonical Partition Function of a one-dimensional fictious gas
(with coordinate position $\beta_i$)
\EQ
\Xi(mr)\,=\,\sum_{N=0}^{\infty} z^N\,Z_N(mr)
\EN
but with a coordinate-dependent activity
\EQ
z_i(mr,\beta_i)\,=\,\frac{1}{2\pi}\,e^{-mr\cosh\beta_i}
\EN
This observation is extremely useful to recover ultraviolet data of the
theory, as the anomalous dimensions of the fields, in terms of massive
quantities \cite{YZ,CMform}.
\end{enumerate}

\noindent
Let us discuss now the main properties of the Form Factors for 2-D Integrable
Massive Field Theories which are the crucial quantities entering the
spectral representation of correlation functions. If not explicit said,
we consider for simplicity the case of a theory with only one self-conjugate
particle. For local scalar operators ${\cal O}(x)$, relativistic invariance
implies that the form factors $F_n$ are functions of the difference of the
rapidities $\beta_{ij}$
\EQ
F_n (\beta_1,\beta_2,\ldots,\beta_n) \,=\,
F_n (\beta_{12},\beta_{13},\ldots,\beta_{ij},\ldots) \,\,\,\,, i<j
\,\,\, .
\EN
Except for the poles corresponding to the one-particle bound states in all
sub-channels, we expect the form factors $F_n$ to be analytic inside the strip
$0 < {\rm Im }\, \be_{ij} < 2\pi$.

The form factors of a hermitian local scalar operator ${\cal O}(x)$ satisfy
a set of equations, known as Watson's equations, which for
integrable systems assume a particularly simple form
\cite{Karowski,Smirnov}
\bea
F_n(\be_1, \dots ,\be_i, \be_{i+1}, \dots, \be_n) &=& F_n(\be_1,\dots,\be_{i+1}
,\be_i ,\dots, \be_n) S(\beta_i-\beta_{i+1}) \,\, ,
\label{permu1}\\
F_n(\be_1+2 \pi i, \dots, \be_{n-1}, \be_n ) &=& F_n(\be_2 ,\ldots,\be_n,
\be_1) = \prod_{i=2}^{n} S(\beta_i-\beta_1) F_n(\be_1, \dots, \be_n)
\,\, .
\nonumber
\eea
In the case $n=2$, eqs.\,(\ref{permu1}) reduce to
\EQ
\begin{array}{ccl}
F_2(\beta)&=&F_2(-\beta)S_2(\beta) \,\, ,\\
F_2(i\pi-\beta)&=&F_2(i\pi+\beta) \,\,\, .
\end{array}
\label{F2}
\EN
It has been shown in \cite{Smirnov} that eqs.\,(\ref{permu1}), together
with the next eqs.\,(\ref{recursive}) and (\ref{respole}), can be regarded
as a system of axioms which defines the whole local operator content of the
theory.

The general solution of Watson's equations can always be brought
into the form \cite{Karowski}
\EQ
F_n(\beta_1,\dots,\beta_n) =K_n(\beta_1,\dots,\beta_n) \prod_{i<j}F_{\rm min}
(\beta_{ij})  \,\, ,
\label{parametrization}
\EN
where $F_{\rm min}(\beta)$ has the properties that it satisfies (\ref{F2}), is
analytic in $0\leq$ Im $\beta\leq \pi$, has no zeros in $0<$ Im $\beta<\pi$,
and converges to a constant value for large values of $\beta$. These
requirements uniquely determine this function, up to a normalization. The
remaining factors $K_n$ then satisfy Watson's equations with $S_2=1$, which
implies that they are completely symmetric, $2 \pi i$-periodic functions of
the $\beta_{i}$. They must contain all the physical poles expected in the form
factor under consideration and must satisfy a correct asymptotic behaviour
for large value of $\beta_i$. Both requirements depend on the nature of the
theory and on the operator $\cal O$.

Notice that one condition on the asymptotic behaviour of the FF
is dictated by relativistic invariance. In fact, a simultaneous shift
in the rapidity variables gives
\EQ
F_n^{\cal O} (\beta_1+\Lambda,\beta_2+\Lambda,\ldots,\beta_n+\Lambda) \,=\,
F_n^{\cal O} (\beta_1,\beta_2,\ldots,\beta_n) \,\, ,
\label{asymp1}
\EN
Secondly, in order to have a power-law bounded ultraviolet behaviour of the
two-point function of the operator ${\cal O}(x)$ (which is the case we
will consider), we have to require that the form factors behave asymptotically
at most as $\exp(k \be_i)$ in the limit $\be_i \rightarrow \infty$, with $k$
being a constant independent of $i$. This means that, once we extract from
$K_n$ the denominator which gives rise to the poles, the remaining part has to
be a symmetric function of the variables $x_i\equiv e^{\beta_i}$, with a finite
number of terms, i.e. a symmetric polynomial in the $x_i$'s.

The pole structure of the form factors induces a set of recursive equations
for the $F_n$ which are of fundamental importance for their explicit
determination. As function of the rapidity differences $\beta_{ij}$, the form
factors $F_n$ possess two kinds of simple poles.

The first kind of singularities (which do not depend on whether or not the
model possesses bound states) arises from kinematical poles located at
$\beta_{ij}=i\pi$. They are related to the one-particle pole in a subchannel
of three-particle states which, in turn, corresponds to a crossing process of
the elastic $S$-matrix. The corresponding residues are computed by the LSZ
reduction \cite{Smirnov} and give rise to a recursive equation
between the $n$-particle and the $(n+2)$-particle form factors
\EQ
-i\lim_{\tilde\beta \rightarrow \beta}
(\tilde\beta - \beta)
F_{n+2}(\tilde\beta+i\pi,\beta,\beta_1,\beta_2,\ldots,\beta_n)=
\left(1-\prod_{i=1}^n S(\beta-\beta_i)\right)\,
F_n(\beta_1,\ldots,\beta_n)  . \label{recursive}
\EN
The second type of poles in the $F_n$ only arise when bound states are present
in the model. These poles are located at the values of $\beta_{ij}$ in the
physical strip which correspond to the resonance angles. Let
$\beta_{ij}=i u_{ij}^k$ be one of such poles associated to the bound state
$A_k$ in the channel $A_i\times A_j$. For the $S$-matrix we have
\EQ
-i\,\lim_{\beta\rightarrow i u_{ij}^k}
(\beta-i u_{ij}^k) \,S_{ij}(\beta)\,=\, \left(\Gamma_{ij}^k\right)^2
\EN
where $\Gamma_{ij}^k$ is the three-particle vertex on mass-shell.
The corresponding residue for the $F_n$ is given by \cite{Smirnov}
\EQ
-i\lim_{\epsilon\rightarrow 0} \epsilon\,
F_{n+1}(\beta+i \overline u_{ik}^j-\epsilon,
\beta-i \overline u_{jk}^i+\epsilon,\beta_1,\ldots,\beta_{n-1})
\,=\,\Gamma_{ij}^k\,F_{n}(\beta,\beta_1,\ldots,\beta_{n-1})
\,\,\, ,
\label{respole}
\EN
where $\overline u_{ab}^c\equiv (\pi-u_{ab}^c)$. This equation establishes
then a recursive structure between the $(n+1)$- and $n$-particle form factors.

Important properties of the FF are pointed out by the following observations:
\begin{enumerate}
\item Notice that the functional and recursive equations satisfied by the Form
Factors do not refer to any operator of the theory! This opens the possibility
to classify the operator content of a massive QFT by computing the
independent solutions of these equations and by associating them to the
corresponding operators, as suggested and investigated in
\cite{CMform,KMform,Christe}. The structure of the local operators in
integrable QFT has been analysed by other points of view in \cite{Smi,Luk,Lec}.
\item The computation of the Form Factors only depend on the $S$-matrix. This
implies that if the $S$-matrix $S(\lambda)$ interpolates between two
(or several) scattering matrices relative to different QFT by varying the
parameter $\lambda$, there should be a corresponding mapping between the
operator content of the QFT encountered in the flow. This correspondance may
be difficult to establish at the perturbative level and therefore it completly
relies on the non-perturbative effects encoded in the exact $S$-matrix. One of
the most striking example relative to this observation is the correspondance
between the Sinh-Gordon model (which is a $Z_2$ invariant model) and the
Bullough-Dodd model (which does not present any symmetry at the perturbative
level) for some particular values of the coupling constants of these models
\cite{FMS}. Another example of the non-perturbative mapping of the
operator content of two QFT has
been established for the Sinh-Gordon and Ising models \cite{ADM}.
\item As it follows from eq.\,(\ref{F2}), if $S(0)=-1$ the two-particle
Form Factor necessarily vanishes at threshold
\EQ
F_{\rm min}(\beta_{ij})\,\simeq\, \beta_{ij}\,\,\,.
\label{smooth}
\EN
This observation is quite important since it permits to understand the
reason of the fast rate of convergence of the spectral series also at short
distance scales, as shown in several significant examples discussed
in the literature \cite{YZ,Camus,YL,Del}. In fact, the correlation
functions are satured by the first matrix elements and for any practical aim,
their computation requires relatively little analytic work. The argument goes
as follows \cite{Camus}. Let us consider the two-point function in the
momentum space
\EQ
G(p)\,\simeq\,\int ds \frac{\sigma(s)}{p^2+s}\,\,\, ,
\EN
where
\EQ
\sigma(s)\,=\,\sum_N\int\frac{\beta_1}{2\pi}\ldots\frac{d\beta_N}{2\pi}
\delta(s-\sum_i^N m\cosh\beta_i)\delta(\sum_i^N m\sinh\beta_i) |F_N|^2
\,\,\,.
\EN
The spectral function $\sigma(s)$ gets more contributions each time that it
passes through a threshold. If the matrix elements $|F_N|^2$ were constants,
its discontinuity at the threshold $(Nm)$ due to the phase space would be
given by
\EQ
\sigma(s)\,\simeq\,\theta(s-(Nm)^2) \left(\sqrt{s-(Nm)^2}\right)^{N-3}\,\,\,.
\EN
However, as consequence of eqs.\,(\ref{parametrization}) and (\ref{smooth})
the spectral function has a much softer behaviour at the different thresholds
\EQ
\sigma(s)\,\simeq\,\theta(s-(Nm)^2) \left(\sqrt{s-(Nm)^2}\right)^{N^2-3}\,\,\,
,
\EN
and therefore the values of the correlation functions are saturated by the
first terms of the spectral representations even at large values of the
momenta.
\end{enumerate}

\resection{One-point Functions in the Ising Model with Boundary}

A relevant aspect of a QFT is the interpolation between its infrared and
ultraviolet regimes. In particular, it is extremely important to establish the
relationship between the most significative parameters associated to
the scaling behaviour in the ultraviolet regime to those which characterize
the infrared properties. The example we choose to illustrate this relationship
is the Ising model with a boundary. The conformal field theory
relative to the scaling behaviour of the fixed point of this model
has been discussed in \cite{CardyB,CL} whereas the breaking of conformal
invariance due to the presence of finite correlation length has been recently
formulated in \cite{GZ}. To compute the scaling dimensions in the presence
of the boundary, it is sufficient to consider the one-point function of the
energy operator $\epsilon_0(t) = <0\mid E(y,t)\mid B>$ and the one-point
function of the disorder operator $\mu_0(t) = <0\mid\mu(y,t)\mid B>$, where
$\mid B >$ is the boundary state (see below). To fix the notation, $t$ is the
distance of the operators from the boundary
whereas $y$ is their parallel coordinate. By translation invariance the
above one-point functions depend only on $t$. In the high temperature phase
(the only one discussed here), these operators share the important property
to couple only to an even number of particles, which we may consider as
massive Majorana fermions described by annihilation and creation operators
$A(\beta)$ and $A^{\dagger}(\beta)$. The mass of the fermion field is
linearly related to the difference of the temperature, $m = 2\pi (T - T_c)$.
The important quantity we need for our computation is the wave function of the
boundary state $\mid B\rangle$ relative to the fixed and free boundary
conditions. This has been determined in \cite{GZ} and its explicit expression
is given by\footnote{There may also be other terms in $\mid B>$ which come
from the bound state. However they do not contribute to the one-point
functions computed in the text.}
\EQ
\mid B > \,=\,\exp\left[\int_0^{\infty}\frac{d\beta}{2\pi}
K(\beta) A^{\dagger}(-\beta) A^{\dagger}(\beta)\right]\mid 0>
\,\,\, .
\label{defstate}
\EN
where
\EQ
K(\beta)\,=\,\left\{\begin{array}{ll}
i \tanh\frac{\beta}{2} & \mbox{fixed b.c.} \\
-i \coth\frac{\beta}{2} & \mbox{free b.c.}
\end{array}
\right.
\EN
The computation of the one-point functions can be done by using the form
factors determined in \cite{Karowski,YZ,CMform}. The energy operator couples
only to the two-particle state,
\EQ
<0\mid E(0)\mid \beta_1,\beta_2> \,=\,2\pi\,i\,m
\sinh\frac{\beta_{12}}{2}
\EN
and its one-point function is given by
\EQ
\epsilon_0(t)\,=\,
i\, m\,\int_0^{\infty}d\beta\,
\sinh\beta\,K(\beta) \,e^{-2mt\cosh\beta} \,\,\, .
\EN
With a simple integration we obtain
\EQ
\frac{\epsilon_0(t)}{m}\,=\,
\left\{\begin{array}{ll}
K_1(2mt) - K_0(2mt) & \mbox{fixed b.c.} \\
-K_1(2mt) - K_0(2mt) & \mbox{free b.c.}
\end{array}
\right.
\EN
($K_0$ and $K_1$ are Bessel functions) and in the ultraviolet limit
($mt \rightarrow 0$) we recover the critical exponent $x=1$ of the
energy operator and the universal amplitudes determined in \cite{CL}.

Concerning the computation of the one-point function of the disorder operator
$\mu(x,t)$, it couples to all states with an even number of particles.
Using the form factors determined in \cite{Karowski,YZ,CMform} and
a simple algebraic identity, the relevant expression can be written in
this case as
\EQ
<0\mid \mu(0,0)\mid -\beta_1,\beta_1,\ldots -\beta_n,\beta_n>\,=\,
\prod_{i=1}^n \tanh\beta_i \times \,
det\,\left(\frac{2\,\sqrt{\cosh\beta_i\cosh\beta_j}}
{\cosh\beta_i+\cosh\beta_j}
\right)\,\,\, .
\EN
The one-point function of $\mu(x,t)$ can be then expressed as a Fredholm
determinant
\EQ
\mu_0(t)\,=\,<0\mid \mu(x,t)\mid B> \,=\,Det
\,(1 - z_{\pm}\,W_{\pm})
 \,\, ,
\label{mu}
\EN
where $z_{\pm} = \pm 1/2\pi$ and $W_{\pm}$ is the kernel of a linear integral
symmetric operator
\begin{eqnarray}
W_{\pm}(\beta_i,\beta_j\mid t) & = &\frac{E_{\pm}(\beta_i, mt)
E_{\pm}(\beta_j,mt)}{\cosh\beta_i + \cosh\beta_j}\,\,\, ;
\label{kernel}\\
E_{\pm}(\beta,mt) & = & e^{-mt\cosh\beta}\,
\sqrt{\cosh\beta \pm 1}
\,\,\, .\nonumber
\end{eqnarray}
The plus sign of the above quantities refers to the fixed b.c. whereas the
minus sign to the free b.c.. In both cases, $\mu_0(t)$ may be written in terms
of the eigenvalues of the integral operator and their multiplicity as
\EQ
\mu_0(t)\,=\,\prod_{i=1}^{\infty} \left(1 - z_{\pm}\,
\lambda_{\pm}^{(i)}\right)^{a_i}
\,\,\, .
\label{eigenvalues}
\EN
As far as $(mt)$ is finite, the kernel is square integrable and therefore all
results valid for bounded symmetric operators apply (see, for instance
\cite{Integral}). However, when $(mt) \rightarrow 0$, the operator becomes
unbounded. The mathematical problem has been studied in the literature
\cite{Proc}: the eigenvalues becomes dense in the interval $(0,\infty)$
according to the distribution
\EQ
\lambda(p)\,=\,\frac{2\pi}{\cosh\pi p} \,\, ,
\EN
whereas, from Mercer's theorem, their multiplicity grows logarithmically as
$a_i \sim \frac{1}{\pi} \ln\frac{1}{mx}$. The critical exponents of the
disorder operator relative to fixed and free boundary conditions are
therefore given by
\EQ
x(z_{\pm})\,= -\,\frac{1}{\pi}\,
\int_0^{\infty} dp \,\ln\left(1 - \frac{2\pi z_{\pm}}{\cosh p}\right)
\,=\,- \frac{1}{8} +\frac{1}{2\pi^2} \arccos^2(-2\pi z_{\pm})\,\,\,.
\EN
Substituting the values of $z_{\pm}$ we obtain the results obtained in
\cite{CL}, i.e. $x=3/8$ for the fixed b.c. and $x=-1/8$ for the free b.c.

\vspace{5mm}
\noindent
{\em Acknowledgements}. I would like to thank C.Ahn, J.L. Cardy, G. Delfino,
A. Fring, A. Koubek and P. Simonetti for their collaboration. I am also
grateful to P. Fendley, H. Saleur and A. Schwimmer for useful discussions.
A special thanks to the organizers of the International Colloquium on
Modern QFT at the Tata Institute in Bombay for their warm hospitality.

\end{document}